\documentclass[9pt,sigconf,screen,natbib=false]{acmart}
\usepackage[T1]{fontenc}
\usepackage[utf8]{inputenc}
\usepackage{microtype}
\usepackage{graphicx}
\usepackage{subcaption}
\usepackage{tabularx}
\usepackage{array}
\newcolumntype{C}{>{\centering\arraybackslash}X}
\usepackage{multirow}
\usepackage{booktabs}
\usepackage{amsmath,amsfonts}
\usepackage{mathtools}
\usepackage{amsthm}
\usepackage{enumitem}
\usepackage{pgfplots}
\usepgfplotslibrary{groupplots}
\pgfplotsset{compat=1.18}
\usepackage{tikz}
\usetikzlibrary{calc,arrows.meta,shapes.geometric,arrows}
\usepackage{bm}
\usepackage{makecell}
\usepackage{xcolor}
\usepackage{colortbl}
\usepackage[most]{tcolorbox}
\usepackage{algorithm}
\usepackage{algorithmic}
\usepackage[capitalize,noabbrev]{cleveref}
\usepackage[textsize=tiny]{todonotes}

\definecolor{myred}{RGB}{227,107,98}
\definecolor{myyellow}{RGB}{255,214,75}
\definecolor{mylight}{RGB}{38,192,159}
\definecolor{yang}{RGB}{29,114,221}
\definecolor{LLMGreen}{HTML}{6F8F5B}
\definecolor{LLMGreenDark}{HTML}{4F6F3E}

\newtcolorbox{llmpromptbox}[1]{%
enhanced,
breakable,
colback=white,
colframe=LLMGreenDark,
arc=7pt,
boxrule=1.1pt,
left=8pt,
right=8pt,
top=10pt,
bottom=10pt,
title=\bfseries\scshape #1,
coltitle=white,
colbacktitle=LLMGreen,
attach boxed title to top left={xshift=10pt,yshift=-2.5mm},
boxed title style={
sharp corners=south,
arc=7pt,
boxrule=0pt,
left=10pt,
right=10pt,
top=2pt,
bottom=2pt,
},
fontupper=\ttfamily\small,
}

\theoremstyle{plain}

\theoremstyle{definition}

\theoremstyle{remark}

\iftrue
\usepackage{titlesec}
\titlespacing*{\section}{0pt}{1.8ex plus .2ex minus .2ex}{0.4ex plus .2ex}
\titlespacing*{\subsection}{0pt}{1.0ex plus .2ex minus .2ex}{0.2ex plus .2ex}
\fi
\graphicspath{{./fig/}}
\usetikzlibrary{matrix}
\renewcommand\footnotetextcopyrightpermission[1]{}

\copyrightyear{2026}
\acmYear{2026}
\setcopyright{none}
\acmConference[MLCAD '26]{ACM/IEEE Workshop on Machine Learning for CAD}{2026}{}
\acmBooktitle{ACM/IEEE Workshop on Machine Learning for CAD (MLCAD '26), 2026}

\begin{document}
\title{Retrieve, Schedule, Reflect: LLM Agents for Chip QoR Optimization}

\author{Yikang Ouyang\hfill Yang Luo\hfill  Dongsheng Zuo\hfill Yuzhe Ma}
\authornote{Corresponding author.}
\affiliation{%
  \institution{The Hong Kong University of Science and Technology (Guangzhou), Guangzhou, China}
  \city{\relax}
  \country{\relax}
}
\renewcommand{\shortauthors}{Ouyang et al.}

\begin{abstract}

Quality-of-results (QoR) optimization through Engineering Change Orders (ECOs), measured by timing, power, and area, is key to chip signoff.
Although commercial EDA tools provide mature optimization algorithms such as gate sizing, threshold-voltage assignment, and buffer insertion, achieving high QoR still heavily depends on iterative human expert intervention.
To address this issue and automate chip ECO, we propose an agentic LLM framework that schedules chip optimizations through direct interaction with EDA tools.
The agents are grounded in natural language expertise expressed as a search tree through retrieval-augmented generation (RAG). 
We further improve scheduling quality with Pareto-driven QoR feedback through language reflection. 
Across eight designs, our framework achieves over 10\% better TNS and 25\% better WNS
than reinforcement learning-based black-box optimization methods.
It also achieves up to 4.5\% lower area and 1.5\% lower power than RL baselines, while running more than 4$\times$ faster. Finally, the agent supports customized tasks expressed in natural language, enabling preferential QoR trade-offs.
Our work is open-sourced at \url{https://github.com/YiKangOY/LLM-PO}.
\end{abstract}

\maketitle

\section{Introduction}

Across modern chip-design flows, ECOs are performed repeatedly from synthesis through signoff to optimize power, performance, and area (QoR).
Commonly used algorithms for improving QoR include gate sizing, threshold-voltage ($V_{th}$) assignment, and buffer insertion/removal~\cite{PD-ICCAD25-Pu,PD-DAC2021-Lu,PD-TODAES23-Lu}.

\begin{figure}[!tbp]
    \centering
       \includegraphics[width=0.88\linewidth]{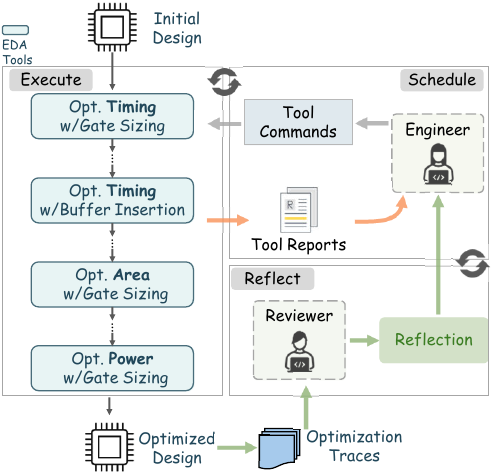}
    \caption{Post-routing ECO optimization as an iterative report-driven schedule--execute--reflect workflow.}
      \label{fig:ExampleECOFlow}
\end{figure}

While these algorithms are integrated into EDA tools and can be invoked easily, achieving competitive QoR still requires extensive human effort.
Due to the NP-hard nature of chip optimization~\cite{PD-TCAD94-Ning}, QoR improvement often saturates and may be trapped in local minima after several iterations~\cite{SYN-TCAD22-Neto}.
As a result, the optimization becomes an iterative process that requires the intervention of expert engineers.
As shown in \Cref{fig:ExampleECOFlow}, engineers inspect reports from EDA tools and previous ECO traces to select the ECO objective and algorithm, which we call \textbf{scheduling}, then invoke EDA tools for execution.
When QoR improvement stagnates, engineers refine the optimization strategy and restart new rounds from the initial design.
Consequently, this \textbf{schedule--execute--reflect} loop is labor-intensive, time-consuming, and demands continuous human monitoring.

Prior works often treat QoR optimization as black-box search over a fixed sequence of EDA tool commands.
These efforts predominantly rely on reinforcement learning since it can optimize over black-box objectives with only numerical QoR feedback~\cite{PD-ICCAD20-Agnesina,PD-DAC23-Lu,PD-MLCAD25-Won,PD-DAC2021-Lu}.
Despite enabling a certain level of automation, these methods face limitations because they primarily rely on numerical QoR feedback.
They cannot comprehend the rich information expressed in natural-language tool reports and logs as human engineers do, leaving critical context for ECO unused and often leading to suboptimal QoR.
Moreover, supporting customized, open-ended optimization tasks remains difficult, since they rely on predefined reward functions~\cite{LLM-NIPS23-Rafailov}.

To address these challenges and enable efficient, high-QoR, and customizable chip optimization, we propose the first agentic large language model (LLM) framework that converts natural-language ECO preferences into iterative EDA tool interactions.
The framework operates in a \textbf{retrieve--schedule--reflect} loop using multiple agents.
First, it \textbf{retrieves} relevant optimization expertise from a library formulated as a search tree, where paths encode human-interpretable ECO strategies described in natural language.
This grounding allows the agent to understand report states and user preferences with expert domain context rather than relying solely on numerical QoR feedback.
Based on the retrieved expertise, the agent then \textbf{schedules} the next objective, strategy, and legal EDA tool options.
Furthermore, to continuously refine decisions under multi-objective QoR trade-offs,
we introduce a Pareto-driven \textbf{reflection} mechanism.
The agent identifies Pareto-optimal optimization traces, summarizes effective preference-aware scheduling strategies, and incorporates these reflections to improve future schedules.

Our contributions are summarized as follows.
\begin{enumerate}
\item We introduce \textbf{Retrieve--Schedule--Reflect}, an LLM multi-agent framework for preference-driven ECO optimization through iterative scheduling of EDA tool usage.
\item We organize expert ECO scheduling knowledge as a natural-language search tree for RAG, grounding scheduling decisions in expert strategies and tool manuals through RAG.
\item We propose Pareto-driven reflection to improve ECO quality by learning from historical timing--power--area trade-offs.
\item On eight designs placed and routed with ASAP7 PDK~\cite{PDK-MICSJ16-Clark}, our framework achieves over 10\% better TNS and 25\% better WNS than RL-based methods. 
It also achieves up to 4.5\% lower area and 1.5\% lower power than them, and delivers higher hypervolume with more than 4$\times$ speedup.
\item Our framework supports customized ECO preferences specified in natural language, enabling controllable trade-offs without task-specific reward engineering.
\end{enumerate}

\section{Preliminaries}
\subsection{ECO and LLM for Chip Design}
Common ECO algorithms include gate sizing~\cite{PD-TCAD94-Ning,PD-ICCAD85-Fishburn}, threshold-voltage assignment, and buffer insertion~\cite{PD-TCAD05-Shi,PD-ISCAS90-Van}.
These algorithms have been integrated into EDA tools~\cite{PD-GOV19-Ajayi, PD-TR26-Synopsys}, but achieving satisfactory QoR still requires engineers to determine objective priorities, algorithm choices, and tool options from reports~\cite{PD-ISQED02-Coudert}.

LLMs have recently been used in chip design as tool-using agents that read reports/logs and propose edits to flow scripts.
For example, ChipNeMo studies domain adaptation for script generation and bug analysis~\cite{LLM-TR23ChipNemo-Liu}, ChatEDA generates scripts to automate RTL-to-GDSII flow~\cite{LLM-TR23ChatEDA-He}, and ORFS-agent tunes RTL-to-GDS flow parameters with an LLM~\cite{PD-MLCAD25-Ghose}.
Different from these works, our framework uses LLMs to understand user-specified QoR preferences and schedule multi-iteration ECO with EDA tool.

\subsection{LLM for Task Scheduling}
Many LLM agents act as task schedulers by interpreting natural-language goals, choosing tools, observing intermediate results, and updating their next decisions.
ReAct is a representative reasoning-action formulation that interleaves explicit LLM reasoning with environment actions, allowing the next-step plan to be updated after each observation~\cite{LLM-TR23React-Yao}.

Two mechanisms are especially useful for improving scheduling quality in iterative preference optimization.
First, \textbf{retrieval-augmented generation (RAG)} lets the agent query external knowledge to guide scheduling with documentation and expert examples~\cite{LLM-NIPS20-Lewis}.
Second, \textbf{reflection} reviews summaries of failures/successes and uses them to refine future schedules, improving behavior in iterative workflows~\cite{LLM-NIPS23-Shinn}.

\subsection{Problem Formulation}
We formulate post-routing ECO as a scheduling and execution process of EDA tool usage under a user-defined natural-language optimization task.
Let $D^{(0)}$ denote the routed design with report $R^{(0)}$ and $\mathcal{P}$ denote the designer-specified task.
At iteration $i$, the scheduler observes $\mathcal{P}$ and accumulated reports $R^{(0:i-1)}$, then selects an optimization objective
\begin{equation}
o^{(i)}\in\mathcal{O}=\{\text{timing},\text{power},\text{area}\},
\end{equation}
and generates strategy guidance for the executor.
The executor converts this guidance into a legal ECO action $A^{(i)}$ with tool-specific options.
The EDA tool applies the action and returns an updated design and report:
\begin{equation}
\big(D^{(i)},R^{(i)}\big)=\mathcal{E}\big(D^{(i-1)},A^{(i)}\big).
\end{equation}

For a completed run with $n$ iterations of ECO, the framework extracts a post-optimization QoR tuple
\begin{equation}
\mathbf{z}=\big(z_{\mathrm{timing}},z_{\mathrm{power}},z_{\mathrm{area}}\big)
\end{equation}
from $R^{(n)}$.
As optimization iterations accumulate, QoR improvements tend to saturate~\cite{SYN-TCAD22-Neto}.
Thus we restart the optimization for multiple rounds, each producing a candidate optimized design.
The final output is the candidate design that best satisfies the natural-language preference $\mathcal{P}$.

\begin{figure}[!t]
    \centering
       \includegraphics[width=.9\linewidth]{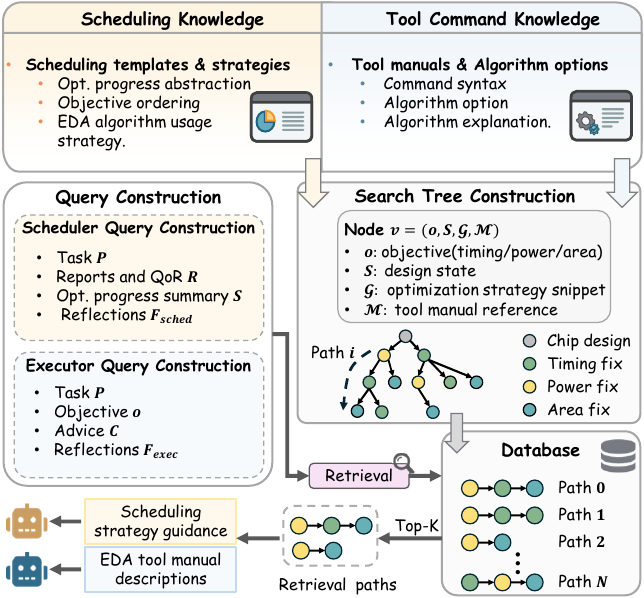}
    \caption{Scheduling strategy constructed as a search tree and path-based RAG.}
      \label{fig:rag}
\end{figure}

\begin{figure*}[!t]
    \centering
       \includegraphics[width=.90\linewidth]{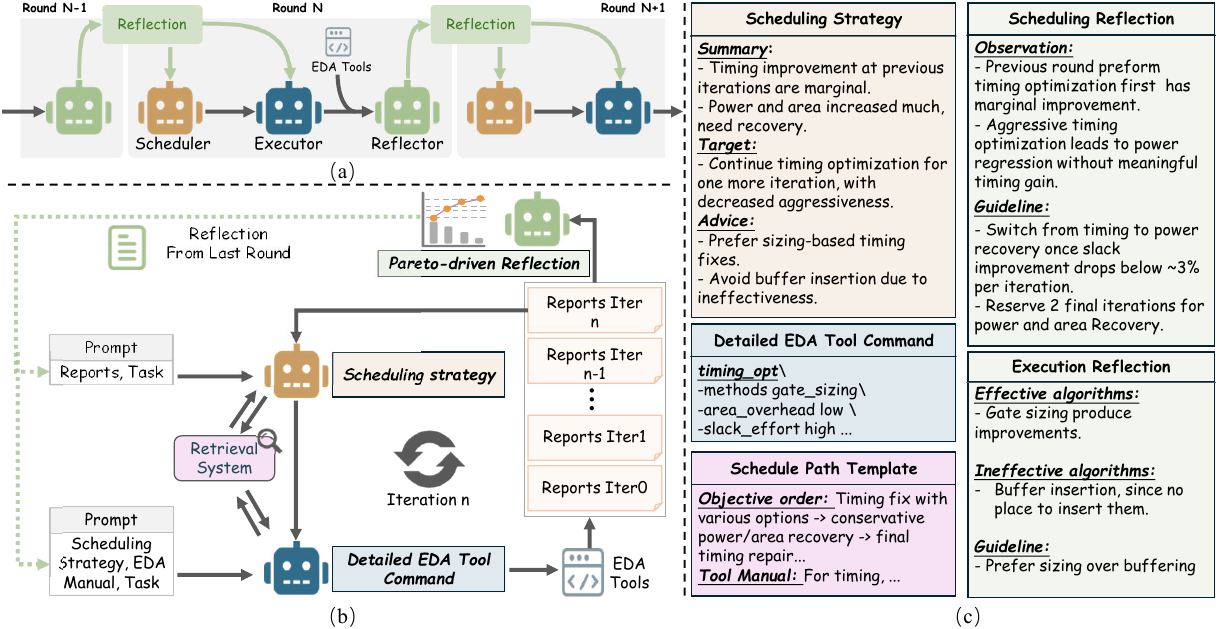}
    \caption{Overview of our framework. (a) Rounds of chip optimization scheduling. (b) Intra-round retrieval, scheduling, and execution. (c) Example prompt and agent response snippets.}
      \label{fig:ECOFlow}
\end{figure*}

\section{Search Tree-based Retrieval-augmented Scheduling}
\label{sec:scheduling_rag}
Effective ECO scheduling requires reusable expert knowledge about objective trade-offs and legal execution compliant to EDA tool manuals.
Retrieval-augmented Generation (RAG) provides a natural mechanism to supply this knowledge at runtime, allowing the agents to condition each decision on relevant optimization strategies and legal tool options.
Inspired by decision-tree formulations for sequential chip optimization~\cite{LS-ICCAD23-Pei}, we model expert schedules as a search tree (ST) and retrieve relevant paths at runtime.

\subsection{Search Tree Representation and Knowledge Store}
The ST is built from papers, reports, expert ECO experiences, and tool manuals, as illustrated in \Cref{fig:rag}.
Each node is associated with an objective $o\in\mathcal{O}=\{\text{timing},\text{power},\text{area}\}$, a natural-language design state $s$, a strategy snippet $\mathcal{G}$ describing how to improve $o$, and objective-specific manual snippets $\mathcal{M}_o$:
Formally, we define a node as
\begin{equation}
  v \triangleq \big(o,\, s,\, \mathcal{G},\, \mathcal{M}_o\big), \qquad o\in\mathcal{O}.
  \label{eq:ST-node}
\end{equation}


An ST path is a variable-length expert scheduling trajectory:
\begin{equation}
  p=(v_1,v_2,\dots,v_L),\quad L\ \text{arbitrary}.
\end{equation}
Paths have no fixed length because satisfying a user preference may require multiple EDA tool calls and objective switches.
One path can represent a complete expert scheduling process and its associated QoR evolution.
It also captures the strategy that designers use to select subsequent objectives and the corresponding algorithms to invoke in EDA tools.
We extract all top-to-leaf paths into a database $\mathcal{D}$ for retrieval.

\subsection{Path-based RAG}
\label{subsec:RAG}
Given a natural-language query $x$ and a candidate ST path $p$, we compute dense embeddings $e_x=\phi(x)$ and $e_p=\phi(p)$ using the same text encoder $\phi$.
We measure relevance using cosine similarity as:
\begin{equation}
  \mathrm{Sim}(x,u)=\frac{e_x\cdot e_u}{\lVert e_x\rVert\,\lVert e_u\rVert}.
\end{equation}
We then retrieve the most relevant expert guidance via top-$k$ similarity search:
\begin{equation}
  \mathrm{TopK}(x,\mathcal{D})=\operatorname*{arg\,top\text{-}k}_{p\in\mathcal{D}}\;\mathrm{Sim}(x,p),
\end{equation}
where $\mathcal{D}$ is the ST path database.
The retrieved paths provide the agent with end-to-end scheduling guidance, which the agent uses to make guided decisions and invoke EDA tools correctly.

\section{Agentic Schedule--Execute--Reflect Framework}
Building upon the ST-based RAG in \Cref{sec:scheduling_rag}, we present an agentic framework for iterative preference-driven QoR optimization as an ECO process.
Starting from a routed design, it performs iterative schedule--execute steps within each round.
During these iterations, RAG provides expert guidance and tool manuals to improve scheduling quality and ensure the executed algorithms are legal for EDA tools.
Once QoR improvement saturates within a round, the framework restarts from the routed design and launches a new round, as shown in \Cref{fig:ECOFlow} (a).
Across rounds, the reflector agent summarizes feedback from historical traces and transfers it to the scheduler and executor to improve scheduling quality in subsequent rounds.


\subsection{Intra-round Scheduling and Execution}
\label{sec:schedule-execute}
We first describe the intra-round scheduling and execution iterations shown in \Cref{fig:ECOFlow}(b).
At iteration $i$ of round $r$, the scheduler interprets the natural-language task, selects the next optimization objective, and forms a high-level strategy.
It observes the accumulated EDA reports and QoR history $R_{r}^{(0:i-1)}$, the user-defined optimization task $\mathcal{P}$, and the previous-round scheduling reflection $\mathcal{F}^{(r-1)}_{\mathrm{sched}}$.

To ground the scheduling in guidance from the ST database $\mathcal{D}$, the scheduler summarizes the user preference, report history, and current QoR state as $s^{(i)}$ and forms a retrieval query:
\begin{equation}
x^{(i)}=\psi_{\mathrm{sched}}\big(\mathcal{P},\,R_{r}^{(0:i-1)},\,s^{(i)},\,i,\,\mathcal{F}^{(r-1)}_{\mathrm{sched}}\big).
\label{eq:scheduleQuery}
\end{equation}
Then the scheduler uses $x^{(i)}$ to retrieve the most relevant expert scheduling paths:
\begin{equation}
\widehat{\mathcal{Y}}^{(i)}=\mathrm{TopK}\big(x^{(i)},\mathcal{D}\big).
\end{equation}
The retrieved paths $\widehat{\mathcal{Y}}^{(i)}$ provide guidance on objective selection and tool-use strategies that match the current design state and requested preference.
Conditioned on $\widehat{\mathcal{Y}}^{(i)}$ and $\mathcal{F}^{(r-1)}_{\mathrm{sched}}$, the scheduler selects the next optimization objective $o^{(i)}\in\mathcal{O}$ and generates strategy guidance $C^{(i)}$.

The executor translates the high-level strategy guidance into an EDA command given $(\mathcal{P},o^{(i)},C^{(i)})$.
To ensure command legality and objective consistency, it queries the tool manual:
\begin{equation}
q^{(i)}=\psi_{\mathrm{exec}}\big(\mathcal{P},\,o^{(i)},\,C^{(i)}\big),
\label{eq:manualQuery}
\end{equation}
and retrieves objective-oriented references:
\begin{equation}
\widehat{\mathcal{M}}_{o^{(i)}}^{(i)}=\mathrm{TopK}\big(q^{(i)},\mathcal{M}_{o^{(i)}}\big).
\end{equation}
Conditioned on the retrieved manual $\widehat{\mathcal{M}}_{o^{(i)}}^{(i)}$ and execution reflection $\mathcal{F}^{(r-1)}_{\mathrm{exec}}$, the executor generates an executable command $A^{(i)}$ that implements $C^{(i)}$ while following the EDA manual and prior legal tool-use syntax~\cite{LLM-NIPS22-Wei}.

Executing $A^{(i)}$ in the EDA tool yields updated design and corresponding reports
\begin{equation}
R_{r}^{(i)}=\mathcal{E}\big(R_{r}^{(i-1)},A^{(i)}\big),
\label{eq:EDACall}
\end{equation}
The updated report is recorded and becomes input to iteration $i+1$.
This intra-round scheduling and execution procedure continues until QoR improvement saturates or the iteration budget is reached.

\begin{algorithm}[!t]
\caption{Iterative ECO scheduling workflow}
\label{alg:overall}
\begin{algorithmic}[1]
\REQUIRE Optimization task $\mathcal{P}$; ST path database $\mathcal{D}$; manuals $\{\mathcal{M}_{o}\}_{o\in\mathcal{O}}$
\STATE Initialize reflections $\mathcal{F}^{(0)}_{\mathrm{sched}},\mathcal{F}^{(0)}_{\mathrm{exec}}$ and trace set $\mathcal{S}\leftarrow\emptyset$
\FOR{$r=1$ to $R_{\max}$}
  \STATE Reset to the initial unoptimized design and obtain initial report $R_r^{(0)}$
  \FOR{$i=1$ to $I_{\max}$}
    \STATE \textbf{Scheduling:} summarize state $s^{(i)}$ and form query $x^{(i)}$ (\Cref{eq:scheduleQuery})
    \STATE Retrieve ST paths; select objective $o^{(i)}$ and strategy guidance $C^{(i)}$
    \STATE \textbf{Execution:} form manual query $q^{(i)}$ and retrieve tool references (\Cref{eq:manualQuery})
    \STATE Generate and execute command $A^{(i)}$ to obtain report $R_{r}^{(i)}$ (\Cref{eq:EDACall})
  \ENDFOR
  \STATE Record trace $S_r$ and terminal QoR vector $\mathbf{z}_r$; update $\mathcal{S}\leftarrow\mathcal{S}\cup\{S_r\}$
  \STATE Filter Pareto-optimal traces from $\mathcal{S}$ and update Pareto-guided reflections (\Cref{sec:reflection})
\ENDFOR
\STATE \textbf{return} optimized design yielded from all rounds of traces under preference $\mathcal{P}$
\end{algorithmic}
\end{algorithm}

\subsection{Inter-round Pareto-guided Reflection}
\label{sec:reflection}
As more optimization operators are applied, QoR improvement often saturates and gets stuck in a local optimum~\cite{PD-ISQED02-Coudert,SYN-TCAD22-Neto}.
To further optimize QoR, our framework draws feedback from previous-round optimization traces  and restarts from the initial design with improved scheduling strategy through reflection.
Pareto dominance provides a multi-objective perspective to evaluate the trade-offs among timing, power, and area across rounds~\cite{PD-TCAD20-Geng,Arch-AAAI24-Bai}.
Thus we maintain a set of Pareto-optimal traces and prompt the reflector to generate feedback for the scheduler and executor in the next round from this set of optimal traces, as shown in \Cref{fig:ECOFlow}.


\subsubsection{Filtering Pareto-optimal Traces}
Let $\mathcal{Z}=\{\mathbf{z}_{1},\mathbf{z}_{2},\dots,\mathbf{z}_{r}\}$ denote the optimized QoR tuple from all completed rounds up to current round $r$:
\begin{equation}
\mathbf{z}_{j}\triangleq \big(z_{j,\mathrm{timing}},\,z_{j,\mathrm{power}},\,z_{j,\mathrm{area}}\big)
\end{equation}
where $\mathbf{z}_{j}$ is extracted from the report of the optimized design at round $j$ (i.e., from $R_{j}^{(n_j)}$).
Larger $z_{j,\mathrm{timing}}$ indicates better timing, while smaller $z_{j,\mathrm{power}}$ and $z_{j,\mathrm{area}}$ indicate better power and area.

Each QoR tuple corresponds to a complete optimization round trace $S_j$ that records executed commands and resulting reports:
\begin{equation}
S_{j}\triangleq \big((A_{j}^{(1)},R_{j}^{(1)}),\,(A_{j}^{(2)},R_{j}^{(2)}),\,\dots,\,(A_{j}^{(n_j)},R_{j}^{(n_j)})\big).
\end{equation}
We compute the Pareto front $\mathcal{Z}_{\mathrm{Pareto}}\subseteq\mathcal{Z}$ under the objectives of maximizing timing and minimizing power and area, then obtain complete optimization traces from the filtered QoR tuples:
\begin{equation}
\mathcal{S}_{\mathrm{Pareto}}=\{S_j \mid \mathbf{z}_j\in \mathcal{Z}_{\mathrm{Pareto}}\}.
\end{equation}

\subsubsection{Scheduler and Executor Reflection}
Optimal QoR is achieved through synergistic work between schedulers and executors.
To benefit both of them through reflection, we prompt the reflector to generate two complementary reflections: a scheduling reflection $\mathcal{F}^{(r)}_{\mathrm{sched}}$ and an execution reflection $\mathcal{F}^{(r)}_{\mathrm{exec}}$.


The scheduler reflection $\mathcal{F}^{(r)}_{\mathrm{sched}}$ extracts patterns of QoR priorities and trade-offs among iterations that lead to Pareto-optimal traces.
It guides when to focus on one objective versus when to pursue others, so that the schedule can satisfy the requested preference.
The executor reflection $\mathcal{F}^{(r)}_{\mathrm{exec}}$ summarizes which EDA commands and option settings are effective, while identifying ineffective choices to avoid.
These reflections are provided to the scheduler and executor in the next round, progressively improving ECO scheduling quality in terms of QoR across rounds.

\begin{table*}[!htbp]
\caption{Chip design statistics and optimization results. 
Here, ``TS'' and ``WS'' denote total slack and worst slack, respectively. 
Ratio (\%) indicates the percentage of the post-ECO metric compared with the original value. 
The best results are \textbf{bolded}.}
\resizebox{\textwidth}{!}{%
\begin{tabular}{c|c|cccccccc|c|c}
\hline
Metric & Method & \begin{tabular}[c]{@{}c@{}}NVDLA\\partition\_m\end{tabular} & \begin{tabular}[c]{@{}c@{}}NVDLA\\partition\_p\end{tabular} & ariane136 & \begin{tabular}[c]{@{}c@{}}mempool\\tile\_wrap\end{tabular} & aes\_256 & hidden1 & hidden2 & hidden5 & Avg. & \makecell{Ratio$\downarrow$ \\(\%)}\\
\hline

\multicolumn{1}{c|}{\multirow{5}{*}{\begin{tabular}[c]{@{}c@{}}TS$\uparrow$\\(ps) \end{tabular}}} & Original & -1.909e+04 & -2.073e+05 & -2.189e+05 & -2.133e+06 & -5.774e+04 & -4.237e+04 & -1.325e+05 & -1.512e+05 & -3.702e+05 & - \\
\multicolumn{1}{c|}{} & Human & -2.901e+03 & -1.174e+04 & \textbf{0.000e+00} & \textbf{0.000e+00} & \textbf{0.000e+00} & \textbf{0.000e+00} & \textbf{0.000e+00} & -4.327e+02 & -1.884e+03 & 0.51\%\\
\multicolumn{1}{c|}{} & RL (A3C) & -8.094e+03 & -1.480e+05 & -1.579e+05 & \textbf{0.000e+00} & -2.979e+03 & -5.374e+03 & -3.587e+02 & -6.237e+02 & -4.042e+04 & 10.92\%\\
\multicolumn{1}{c|}{} & RL (PPO) & -1.611e+03 & -5.022e+03 & \textbf{0.000e+00} & -1.558e+06 & \textbf{0.000e+00} & -2.547e+03 & -1.010e+03 & -4.289e+02 & -1.961e+05 & 52.96\%\\
\multicolumn{1}{c|}{} & \textbf{Ours} & \textbf{-1.588e+03} & \textbf{-4.739e+03} & \textbf{0.000e+00} & \textbf{0.000e+00} & \textbf{0.000e+00} & \textbf{0.000e+00} & \textbf{0.000e+00} & \textbf{-4.259e+02} & \textbf{-8.441e+02} & \textbf{0.23\%}\\
\hline

\multirow{5}{*}{\begin{tabular}[c]{@{}c@{}}WS$\uparrow$\\(ps)\end{tabular}} & Original & -3.657e+02 & -2.784e+02 & -1.667e+02 & -4.777e+02 & -5.820e+01 & -1.314e+02 & -1.823e+02 & -8.691e+01 & -2.184e+02 & -\\
 & Human & -9.276e+01 & -6.650e+01 & \textbf{3.431e-01} & 4.700e-01 & \textbf{2.400e+00} & \textbf{8.670e+01} & \textbf{2.330e-01} & \textbf{-3.117e+01} & \textbf{-1.254e+01} & \textbf{5.74\%}\\
 & RL (A3C) & -2.144e+02 & -2.601e+02 & -1.935e+02 & 2.761e+00 & -3.310e+01 & -1.225e+02 & -1.573e+01 & \textbf{-3.117e+01} & -1.085e+02 & 49.68\%\\
 & RL (PPO) & \textbf{-6.468e+01} & -4.759e+01 & 2.182e-01 & -3.101e+02 & 2.400e-03 & -8.904e+01 & -2.912e+01 & \textbf{-3.117e+01} & -7.144e+01 & 32.71\%\\
 & \textbf{Ours} & \textbf{-6.468e+01} & \textbf{-4.743e+01} & 1.660e-01 & \textbf{3.700e+01} & 8.000e-04 & 3.400e-03 & 3.960e-02 & \textbf{-3.117e+01} & -1.326e+01 & 6.07\%\\
 \hline

\multirow{5}{*}{\makecell{Area$\downarrow$\\($\mu m^2$)}} & Original & 2.728e+03 & 9.570e+03 & 3.266e+04 & 2.813e+04 & 2.728e+04 & 5.355e+03 & 3.307e+04 & 2.734e+04 & 2.077e+04 & - \\
 & Human & 2.879e+03 & 9.655e+03 & 2.938e+04 & 2.728e+04 & 2.730e+04 & 5.362e+03 & 2.968e+04 & 2.753e+04 & 1.988e+04 & 95.71\%\\
 & RL (A3C) & \textbf{2.757e+03} & \textbf{9.626e+03} & 3.253e+04 & 2.831e+04 & 2.725e+04 & 5.353e+03 & 3.307e+04 & \textbf{2.712e+04} & 2.075e+04 & 99.90\%\\
 & RL (PPO) & 2.883e+03 & 9.653e+03 & 3.267e+04 & \textbf{2.720e+04} & 2.731e+04 & 5.370e+03 & 3.296e+04 & 2.744e+04 & 2.068e+04 & 99.61\%\\
 & \textbf{Ours} & 2.898e+03 & 9.665e+03 & \textbf{2.937e+04} & 2.735e+04 & \textbf{2.692e+04} & \textbf{5.342e+03} & \textbf{2.947e+04} & 2.753e+04& \textbf{1.982e+04} & \textbf{95.43\%}\\
 \hline

\multirow{5}{*}{\makecell{Power$\downarrow$\\(W)}} & Original & 5.250e-03 & 5.450e-02 & 5.928e-01 & 4.640e-02 & 3.037e-01 & 2.200e-02 & 5.863e-01 & 4.440e-01 & 2.569e-01 & - \\
 & Human & 5.710e-03 & 5.680e-02 & \textbf{5.841e-01} & 4.770e-02 & 3.056e-01 & 2.210e-02 & 5.752e-01 & \textbf{4.544e-01} & 2.565e-01 & 99.84\%\\
 & RL (A3C) & \textbf{5.518e-03} & \textbf{5.580e-02} & 5.928e-01 & 4.900e-02 & 3.110e-01 & \textbf{2.200e-02} & 5.864e-01 & 4.550e-01 & 2.597e-01 & 101.09\%\\
 & RL (PPO) & 5.866e-03 & 5.700e-02 & 5.929e-01 & 4.790e-02 & 3.056e-01 & 2.220e-02 & 5.861e-01 & 4.553e-01 & 2.591e-01 & 100.87\%\\
 & \textbf{Ours} & 5.875e-03 & 5.700e-02 & 5.842e-01 & \textbf{4.700e-02} & \textbf{3.006e-01} & \textbf{2.200e-02} & \textbf{5.747e-01} & 4.557e-01 & \textbf{2.559e-01} & \textbf{99.61\%}\\
 \hline
\end{tabular}
}
\vspace{-10pt}
\label{tab:asap7-stats}
\end{table*}

\section{Experimental Results}
\subsection{Experimental Setup}
To evaluate the generality of our approach, we use open-source chip designs implemented in the ASAP7 PDK~\cite{PDK-MICSJ16-Clark} from the ICCAD'24 dataset~\cite{PD-ICCAD24-Wu}.
We select 8 designs from this benchmark, whose statistics are provided in \Cref{sec:designStats}.
We evaluate three core QoR metrics: timing (both WNS and TNS), power, and area.
All optimizations and report generation are performed using \textit{Synopsys PrimeECO}, a widely used signoff tool.
All experiments are conducted on a machine with two AMD EPYC 7543 32-core CPUs and an NVIDIA RTX 3090 GPU.
Unless otherwise specified, we use \textbf{GPT-5-mini} (2025-08-07) as the default backbone LLM for the agents.

To the best of our knowledge, there is no established baseline that directly targets iterative ECO.
We therefore adapt two reinforcement learning (RL) baselines with A3C for architecture search~\cite{Arch-AAAI24-Bai} and a PPO algorithm for EDA tool parameter search~\cite{PD-TODAES26-Hsiao}, and train them for 50 episodes (rounds), each with 10 iterations, to search fixed-length sequences of EDA commands from QoR feedback.
The training and hyperparameter details are in the \Cref{subsec:RL-setup}.
In addition, we include a human baseline by asking an experienced engineer to operate EDA tools for 10 rounds, each with 10 iterations.
Our framework runs with the same configuration as the human baseline.

\subsection{Optimization Results}
As a common practice, timing slack is treated as a hard constraint and is prioritized first, since negative slack may lead to functional failure of the design.
Once timing violations are resolved, power and area are optimized as secondary objectives under the satisfied timing constraint.
Our primary optimization preference is thus specified in natural language: \textbf{achieve non-negative worst slack (WNS), and once this timing target is reached, minimize the normalized area--power cost, computed as $\mathbf{Area}/\mathbf{Area}_{\mathbf{init}} + \mathbf{Power}/\mathbf{Power}_{\mathbf{init}}$}.
This preference is also used to select the optimized design to present from all launched optimization rounds (or episodes for RL baselines).
\Cref{tab:asap7-stats} compares our method with RL baselines and manual scheduling by an experienced engineer.

Our framework achieves the strongest average timing optimization.
For TNS, it improves the original timing violation by 99.77\%, compared with 89.08\%/47.04\% for A3C~\cite{Arch-AAAI24-Bai}/PPO~\cite{PD-TODAES26-Hsiao}, which corresponds to 10.7\%/52.7\% better TNS than A3C/PPO.
For WNS, it reduces the average violation from $-218.4$ ps to $-13.26$ ps, which corresponds to 43.6\%/26.6\% better WNS than A3C/PPO, while remaining comparable to the human baseline ($-12.54$ ps).
On designs with remaining negative slack after optimization, such as NVDLA\_partition\_m/p and hidden5, our method still achieves the least negative slack.

In addition to optimizing timing, our framework also preserves power and area.
It achieves the lowest average area among all optimized results (95.43\% of the original design) and the lowest average power (99.61\% of the original design).
The slightly larger overheads on the NVDLA\_partition\_m/p reflect the cost of aggressive timing repair for severe violations.
Overall, the QoR results under a common preference that prioritizes timing before power and area demonstrate that our agentic framework is stronger for iterative ECO optimization than RL-based black-box search, even though RL observes more EDA feedback traces during training.

We further present the hypervolume progression during the optimization process of the Pareto frontier over WNS, power, and area for selected design cases in \Cref{fig:pareto-hv-metrics}.
We set the area and power reference points to $1.1\times$ their initial values to avoid zero hypervolume since optimizing timing may slightly degrade power and area.
Our agentic framework reaches a larger hypervolume within 10 rounds, whereas PPO~\cite{PD-TODAES26-Hsiao} and A3C~\cite{Arch-AAAI24-Bai} require many more training episodes and still converge to a lower hypervolume.
This result indicates that our framework not only satisfies the default timing-first preference, but also explores a stronger Pareto frontier for comprehensive QoR trade-offs.

\begin{figure*}[t]
  \centering
  \begin{tikzpicture}
\begin{groupplot}[
    group style={
        group size=4 by 1,
        horizontal sep=1.25cm,
    },
    width=0.225\textwidth,
    height=0.16\textwidth,
    xmin=0,
    xmax=50,
    xtick={0,10,20,30,40,50},
    xlabel={Rounds},
    grid=both,
    grid style={line width=.1pt, draw=gray!20},
    major grid style={line width=.2pt, draw=gray!35},
    tick label style={font=\normalsize},
    label style={font=\normalsize},
    title style={font=\normalsize, yshift=-4pt},
    scaled y ticks=false,
    y tick label style={
        /pgf/number format/sci,
        /pgf/number format/precision=1,
    },
]

\nextgroupplot[
    title={mempool\_tile\_wrap},
    legend to name=paretohvlegend,
    ylabel={Hypervolume},
]
\addplot+[mylight, thick, mark=none, dashed] table [x=episode, y=hv1_2_3, col sep=comma] {pgfplots/pareto_hv_mempool_tile_wrap_ppo.csv};
\addlegendentry{PPO~\cite{PD-TODAES26-Hsiao}}
\addplot+[yang, thick, mark=none, dashed] table [x=episode, y=hv1_2_3, col sep=comma] {pgfplots/pareto_hv_mempool_tile_wrap_a3c.csv};
\addlegendentry{A3C~\cite{Arch-AAAI24-Bai}}
\addplot+[black, thick, mark=none] table [x=iteration_index, y=hv1_2_3, col sep=comma] {pgfplots/pareto_hv_mempool_tile_wrap_agent.csv};
\addlegendentry{Ours}

\nextgroupplot[
    title={aes\_256},
]
\addplot+[mylight, thick, mark=none, dashed] table [x=episode, y=hv1_2_3, col sep=comma] {pgfplots/pareto_hv_aes_256_ppo.csv};
\addplot+[yang, thick, mark=none, dashed] table [x=episode, y=hv1_2_3, col sep=comma] {pgfplots/pareto_hv_aes_256_a3c.csv};
\addplot+[black, thick, mark=none] table [x=iteration_index, y=hv1_2_3, col sep=comma] {pgfplots/pareto_hv_aes_256_agent.csv};

\nextgroupplot[
    title={NVDLA\_partition\_p},
]
\addplot+[mylight, thick, mark=none, dashed] table [x=episode, y=hv1_2_3, col sep=comma] {pgfplots/pareto_hv_NVDLA_partition_p_ppo.csv};
\addplot+[yang, thick, mark=none, dashed] table [x=episode, y=hv1_2_3, col sep=comma] {pgfplots/pareto_hv_NVDLA_partition_p_a3c.csv};
\addplot+[black, thick, mark=none] table [x=iteration_index, y=hv1_2_3, col sep=comma] {pgfplots/pareto_hv_NVDLA_partition_p_agent.csv};

\nextgroupplot[
    title={hidden1},
]
\addplot+[mylight, thick, mark=none, dashed] table [x=episode, y=hv1_2_3, col sep=comma] {pgfplots/pareto_hv_hidden1_ppo.csv};
\addplot+[yang, thick, mark=none, dashed] table [x=episode, y=hv1_2_3, col sep=comma] {pgfplots/pareto_hv_hidden1_a3c.csv};
\addplot+[black, thick, mark=none] table [x=iteration_index, y=hv1_2_3, col sep=comma] {pgfplots/pareto_hv_hidden1_agent.csv};

\end{groupplot}
\node[anchor=west, font=\small] at ($(group c4r1.east)+(0.18cm,0)$) {\pgfplotslegendfromname{paretohvlegend}};
\end{tikzpicture}
  \caption{Hypervolume progression for selected design cases. PPO and A3C are plotted over 50 training episodes, while our agent is plotted over 10 rounds.}
  \label{fig:pareto-hv-metrics}
\end{figure*}

\begin{table}[!tbp]
  \centering
  \caption{Customized optimization with allowable WNS degradation of 5\% of the clock period. The power and area reductions are compared with \textbf{Ours} in \Cref{tab:asap7-stats}. Additional power-breakdown comparisons are in \Cref{sec:power-breakdown}.}
  \label{tab:user-preference}
  \resizebox{0.96\columnwidth}{!}{%
  \begin{tabular}{lcccc}
    \toprule
    Design &
    \makecell{Allowable \\ WS (ps)} &
    \makecell{Achieved \\ WS (ps)} &
    \makecell{Power \\ Reduction} &
    \makecell{Area \\ Reduction} \\
    \midrule
    ariane136 & -35   & -34.90 & 0.05\% & 0.35\% \\
        \hline

    \makecell[l]{mempool\\tile\_wrap} & -130 & -130.00    & 1.90\% & 0.21\% \\
        \hline

    aes\_256  & -27.5 & -27.49  & 1.34\% & 1.85\% \\
        \hline

    hidden1   & -30   & -14.23 & 0.00\% & 0.11\% \\
        \hline

    hidden2   & -35   & -34.29  & 0.00\% & 0.06\% \\
    
    \bottomrule
  \end{tabular}
  }
\end{table}

\subsection{Customized Optimization with User Preference}
In practice, engineers often optimize under user-specified preferences, such as trading limited timing margin for lower power or area~\cite{PD-ISQED02-Coudert,PD-VLSISOC20-Sharma}.
Such preference-driven objectives are difficult to realize with black-box optimizers (e.g., RL) without careful reward engineering~\cite{LLM-NIPS23-Rafailov}.
In contrast, LLM-based agents can directly interpret natural-language preferences and adjust scheduling accordingly.

We select five designs from \Cref{tab:asap7-stats} that achieve timing-clean solutions under the default preference, and then relax timing by allowing WNS degradation up to 5\% of the clock period.
\Cref{tab:user-preference} reports the allowable/achieved worst slack and the power/area reductions relative to \textbf{Ours} in \Cref{tab:asap7-stats}.

Across all five designs, the achieved worst slacks satisfy the allowable limits.
Power reductions are near-zero on hidden1 and hidden2 because clock power dominates total power (see \Cref{tab:asap7-stats-original}), limiting overall reduction.
When excluding clock power, the customized objective yields more substantial reductions, as analyzed in \Cref{sec:power-breakdown}.
Overall, these results show that our framework can follow language-instructed preferences and satisfies customized QoR objectives during ECO.

\begin{table}[t]
  \centering
  \caption{Ablation results averaged over all designs in \Cref{tab:asap7-stats}. The best results are \textbf{bolded}.}
  \label{tab:ablation}
  \resizebox{\columnwidth}{!}{%
  \begin{tabular}{lcccc}
    \toprule
    Method &
    \makecell{Total\\Slack (ps)} &
    \makecell{Worst\\Slack (ps)} &
    Power (W) &
    Area ($\mu m^2$) \\
    \midrule
    \textbf{GPT-5-mini}
      & \textbf{-8.441e+02} & \textbf{-1.326e+01} & 2.559e-01 & 1.981e+04 \\
          \hline

    GPT-4o-mini
      & -1.534e+03 & -2.536e+01 & 2.581e-01 & 2.039e+04 \\
          \hline

    DeepSeek-V3.2
      &  -1.670e+03 & -2.195e+01 & \textbf{2.557e-01} & \textbf{1.977e+04} \\
          \hline

      w/o RAG, w/ Reflect
      &  -1.559e+03 & -2.668e+01 & 2.581e-01 & 1.992e+04 \\
          \hline

     w/ RAG, w/o Reflect
      & -1.829e+03 & -2.371e+01 & 2.558e-01 & 1.986e+04 \\
          \hline

      w/o RAG, w/o Reflect
      & -1.973e+04 & -2.716e+01 & 2.566e-01 & 2.005e+04 \\
    \bottomrule
  \end{tabular}
  }
\end{table}

\subsection{Ablation Study}

We perform ablation studies on the RAG and reflection modules of our framework and test the framework with different backbone LLMs.
Results averaged over all designs are reported in \Cref{tab:ablation}.
Replacing the default GPT-5-mini with GPT-4o-mini (2024-07-18) yields a modest degradation in both slack metrics.
We also use open-source DeepSeek-V3.2 (2025-12-01)~\cite{LLM-TR25-DSLiu}, which improves power and area slightly, but results in slightly worse timing slack.

Removing the RAG and reflection mechanisms individually results in QoR degradation, with the removal of both leading to the worst results.
Thus, both natural-language expertise grounding and QoR-aware reflection are essential for a successful ECO.

\subsection{Runtime and Cost Analysis}

We report average token usage, API cost, and runtime in \Cref{tab:runtime-cost}.
The RL baselines run for over 50  episodes (rounds) but still achieve lower QoR than our method with 10 rounds.
We therefore directly compare the end-to-end runtime of RL over 50 episodes with our framework over 10 rounds.
Under this setting, our full method finishes in 3.54h, while training RL takes 15.38h, giving a \(4.3\times\) runtime reduction.
This speed advantage comes in addition to the QoR gains reported above, showing that language-guided iterative optimization is more efficient than black-box RL for ECO flows.

Removing RAG and Reflection components reduces token usage and runtime, but also weakens QoR.
Without RAG, the average token count drops from 1.45M to 0.58M and runtime decreases from 3.54h to 1.55h; removing both RAG and reflection further reduces the cost to 0.54M tokens and 1.48h.
Even with these components, the full framework remains much faster than RL and incurs modest token cost, making it practical for ECO flows.

\Cref{fig:runtime-breakdown} shows that the remaining runtime is largely due to EDA tool execution, especially on large designs.
For the larger NVDLA case, EDA calls account for 73.6\% of total runtime, compared with 17.9\% for LLM calls and 8.5\% for other overhead.
Thus, for industrial large designs, LLM interaction is mostly amortized by expensive EDA runs, and the agent acts as a preference-aware scheduling layer rather than the dominant runtime bottleneck.

\begin{table}[t]
  \centering
  \caption{Runtime and cost analysis across methods averaged over all designs.}
  \label{tab:runtime-cost}


  \resizebox{0.9\columnwidth}{!}{%
  \begin{tabular}{lccc}
    \toprule
    Method & Token & Cost (\$) & Runtime (h) \\
    \midrule
    A3C~\cite{Arch-AAAI24-Bai}/PPO~\cite{PD-TODAES26-Hsiao}
      & --
      & --
      & 15.38/17.45 \\ \hline
    \textbf{Ours}
      & 1{,}451{,}290
      & 1.25
      & 3.54 \\ \hline
    Ours (GPT-4o-mini)
      & 911{,}470
      & 0.16
      & 0.93 \\ \hline
    Ours (DeepSeek-V3.2)
      & 1{,}003{,}061
      & 1.16
      & 2.61 \\ \hline
    Ours w/o RAG, w/ Reflect
      & 582{,}528
      & 0.46
      & 1.55 \\ \hline
    Ours w/ RAG, w/o Reflect
      & 1{,}393{,}026
      & 1.18
      & 2.83 \\ \hline
    Ours w/o RAG, w/o Reflect
      & 544{,}610
      & 0.44
      & 1.48 \\
    \bottomrule
  \end{tabular}
  }
\end{table}

\begin{figure}[t]
  \centering
  \usetikzlibrary{arrows.meta}

\def\innerradius{0\textwidth}
\def\outerradius{0.04\textwidth}

\definecolor{A0}{HTML}{A93226}
\definecolor{B0}{HTML}{76448A}
\definecolor{C0}{HTML}{1F618D}
\definecolor{D0}{HTML}{148F77}
\definecolor{E0}{HTML}{9A7D0A}
\definecolor{F0}{HTML}{F4D03F}
\definecolor{G0}{HTML}{E67E22}
\definecolor{H0}{HTML}{7B7D7D}
\definecolor{I0}{HTML}{FCE4EC}
\definecolor{J0}{HTML}{1C2833}
\definecolor{myred}{RGB}{227,107,98}
\definecolor{myyellow}{RGB}{255,214,75}
\definecolor{mymiddleblue}{RGB}{139,201,246} 
\definecolor{mylight}{RGB}{38,192,159}

\newcommand{\wheelchartwithlegend}[1]{
  \pgfmathsetmacro{\totalnum}{0}
  \foreach \value/\colour/\name in {#1} {
      \pgfmathparse{\value+\totalnum}
      \global\let\totalnum=\pgfmathresult
  }

    \begin{tikzpicture}

    \pgfmathsetmacro{\wheelwidth}{\outerradius-\innerradius}
    \pgfmathsetmacro{\midradius}{(\outerradius+\innerradius)/2}

    \begin{scope}[rotate=90]

    \coordinate (L-0) at (\outerradius-7mm,-\outerradius-1cm);

    \pgfmathsetmacro{\cumnum}{0}
    \foreach [count=\i,remember=\i as \j (initially 0)] \value/\colour/\name in {#1} {
          \pgfmathsetmacro{\newcumnum}{\cumnum + \value/\totalnum*360}

          \pgfmathsetmacro{\percentage}{\value/\totalnum*100}
          \pgfmathsetmacro{\midangle}{-(\cumnum+\newcumnum)/2}

          \pgfmathparse{
             (-\midangle<180?"west":"east")
          } \edef\textanchor{\pgfmathresult}
          \pgfmathsetmacro\labelshiftdir{1-2*(-\midangle>180)}

          \fill[\colour] (-\cumnum:\outerradius) arc (-\cumnum:-(\newcumnum):\outerradius) --
          (-\newcumnum:\innerradius) arc (-\newcumnum:-(\cumnum):\innerradius) -- cycle;

          \draw  [Circle-,thin] node [append after command={(\midangle:\midradius pt) -- (\midangle:\outerradius + 1ex) -- (\tikzlastnode)}] at (\midangle:\outerradius + 1ex) [xshift=\labelshiftdir*0.5cm,inner sep=0pt, outer sep=0pt, ,anchor=\textanchor]{\pgfmathprintnumber{\percentage}\thinspace\%};

          \node [anchor=north west,text width=3cm,font=\footnotesize] (L-\i) at (L-\j.south west) {\name};
          \fill [fill=\colour] ([xshift=-3pt,yshift=1mm]L-\i.north west) rectangle ++(-2mm,5mm);

          \global\let\cumnum=\newcumnum
      }
    \end{scope}
  \end{tikzpicture}
} 

\begin{tikzpicture}
  \begin{scope}
    \pgfmathsetmacro{\totalnum}{0}
    \foreach \value/\colour/\name in {
      1235.954/mylight/{LLM},
      5097.444/myyellow/{EDA},
      588.804/myred/{Other}
    }{
      \pgfmathparse{\value+\totalnum}
      \global\let\totalnum=\pgfmathresult
    }

    \pgfmathsetmacro{\wheelwidth}{\outerradius-\innerradius}
    \pgfmathsetmacro{\midradius}{(\outerradius+\innerradius)/2}

    \begin{scope}[rotate=90]
      \coordinate (L-0) at (\outerradius-7mm,-\outerradius-1cm);

      \pgfmathsetmacro{\cumnum}{0}
      \foreach [count=\i,remember=\i as \j (initially 0)] \value/\colour/\name in {
      1235.954/mylight/{LLM},
      5097.444/myyellow/{EDA},
      588.804/myred/{Other}
      }{
        \pgfmathsetmacro{\newcumnum}{\cumnum + \value/\totalnum*360}

        \pgfmathsetmacro{\percentage}{\value/\totalnum*100}
        \pgfmathsetmacro{\midangle}{-(\cumnum+\newcumnum)/2}

        \pgfmathparse{(-\midangle<180?"west":"east")} \edef\textanchor{\pgfmathresult}
        \pgfmathsetmacro\labelshiftdir{1-2*(-\midangle>180)}

        \fill[\colour] (-\cumnum:\outerradius) arc (-\cumnum:-(\newcumnum):\outerradius) --
        (-\newcumnum:\innerradius) arc (-\newcumnum:-(\cumnum):\innerradius) -- cycle;

        \draw [Circle-,thin] node [append after command={(\midangle:\midradius pt) -- (\midangle:\outerradius + 1ex) -- (\tikzlastnode)}]
          at (\midangle:\outerradius + 1ex)
          [xshift=\labelshiftdir*0.2cm,inner sep=0pt, outer sep=0pt, anchor=\textanchor]
          {\pgfmathprintnumber{\percentage}\thinspace\%};


        \global\let\cumnum=\newcumnum
      }
    \end{scope}
    \node[font=\footnotesize] at (0,-\outerradius-6mm) {NVDLA\_partition\_m: 3.45h};
  \end{scope}

  \begin{scope}[xshift=0.25\textwidth]
    \pgfmathsetmacro{\totalnum}{0}
    \foreach \value/\colour/\name in {
      6285.327/mylight/{LLM},
      4440.000/myyellow/{EDA},
      2021.676/myred/{Other}
    }{
      \pgfmathparse{\value+\totalnum}
      \global\let\totalnum=\pgfmathresult
    }

    \pgfmathsetmacro{\wheelwidth}{\outerradius-\innerradius}
    \pgfmathsetmacro{\midradius}{(\outerradius+\innerradius)/2}

    \begin{scope}[rotate=90]
      \pgfmathsetmacro{\cumnum}{0}
      \foreach [count=\i,remember=\i as \j (initially 0)] \value/\colour/\name in {
      6285.327/mylight/{LLM},
      4440.000/myyellow/{EDA},
      2021.676/myred/{Other}
      }{
        \pgfmathsetmacro{\newcumnum}{\cumnum + \value/\totalnum*360}

        \pgfmathsetmacro{\percentage}{\value/\totalnum*100}
        \pgfmathsetmacro{\midangle}{-(\cumnum+\newcumnum)/2}

        \pgfmathparse{(-\midangle<180?"west":"east")} \edef\textanchor{\pgfmathresult}
        \pgfmathsetmacro\labelshiftdir{1-2*(-\midangle>180)}

        \fill[\colour] (-\cumnum:\outerradius) arc (-\cumnum:-(\newcumnum):\outerradius) --
        (-\newcumnum:\innerradius) arc (-\newcumnum:-(\cumnum):\innerradius) -- cycle;

        \draw [Circle-,thin] node [append after command={(\midangle:\midradius pt) -- (\midangle:\outerradius + 1ex) -- (\tikzlastnode)}]
          at (\midangle:\outerradius + 1ex)
          [xshift=\labelshiftdir*0.1cm,inner sep=0pt, outer sep=0pt, anchor=\textanchor]
          {\pgfmathprintnumber{\percentage}\thinspace\%};


        \global\let\cumnum=\newcumnum
      }
    \end{scope}
    \node[font=\footnotesize] at (0,-\outerradius-6mm) { hidden1: 1.89h };

    \def\legendx{0.08\textwidth}
    \coordinate (LegendCenter) at (-0.125\textwidth,\outerradius+5mm);
    \coordinate (LegendBase) at ($(LegendCenter)+(-\legendx,0)$);
    \foreach [count=\i] \value/\colour/\name in {
      6285.327/mylight/{LLM},
      4440.000/myyellow/{EDA},
      2021.676/myred/{Other}
    }{
      \pgfmathtruncatemacro{\col}{\i-1}
      \path (LegendBase) ++(\legendx*\col,0) coordinate (Legend-\i);
      \node [anchor=west,font=\footnotesize] at (Legend-\i) {\name};
      \fill [fill=\colour] ([xshift=-1mm,yshift=-1mm]Legend-\i) rectangle ++(-2mm,2mm);
    }
    \draw[thin,rounded corners=1pt]
      ([xshift=-5.3mm,yshift=2mm]LegendBase)
      rectangle
      ([xshift=0.22\textwidth,yshift=-2mm]LegendBase);
  \end{scope}
\end{tikzpicture}
  \caption{Runtime breakdown for NVDLA\_partition\_m and
  hidden1.
  The \textit{Other} sector denotes other runtime costs like EDA tool report parsing.}
  \label{fig:runtime-breakdown}
\end{figure}

\section{Conclusion}
This paper presents \textbf{Retrieve--Schedule--Reflect}, an agentic LLM framework that improves chip QoR by scheduling iterative ECO actions through EDA tools.
Our framework uses search-tree retrieval to ground scheduling decisions in natural-language expert knowledge, executes ECO actions through EDA tools, and applies Pareto-driven reflection to improve timing--power--area trade-offs across rounds.
Experiments on eight designs show that our method achieves stronger timing optimization than reinforcement-learning baselines while reducing power and area, with more than $4\times$ runtime speedup.
The framework also supports user-defined optimization preferences expressed in natural language, enabling customized trade-offs without task-specific reward design.
\bibliographystyle{IEEEtran}
\bibliography{ref/Top-sim,ref/PD,ref/LLM,ref/additional}

\newpage
\appendix
\onecolumn
\section{Scheduling Search Tree}
\label{sec:ST}
Here we discuss the contents of the search-tree nodes and paths used for retrieval.
Each node $v$ in the search tree contains a corresponding objective $o$ among power, timing, and area.
It also contains a high-level natural-language design summary $s$, such as ``the design suffers from severe timing violation of x\% of the clock period'' or ``power has increased by x\% compared with the initial design.''
The node further stores a language-instructed strategy $\mathcal{G}$ for using EDA tools, such as ``explore various gate-sizing options'' or ``recover power aggressively while tolerating limited negative slack.''
A segment of objective-specific tool manual is also associated with each node to guide legal EDA tool use; we omit the licensed manual content.

The tree is built by summarizing academic papers, technical reports, expert ECO strategies, and tool manuals, and is filtered by engineers to ensure validity for chip optimization.
For example, a node with summary $s_i$ stating that ``timing violation is small'' should not transition to a next node $v_{i+1}$ whose summary $s_{i+1}$ states that ``timing violation is significant'' after timing optimization is performed.

By extracting nodes in the search tree into paths, we obtain natural-language scheduling templates that describe possible design-state transitions, objective preferences, and corresponding strategies.
An example path is shown below.
It summarizes design-state changes and objective choices in natural language, and it describes high-level EDA tool usage such as exploring different options or increasing effort.
The scheduler performs RAG at each schedule step to retrieve paths matched to the current design state, QoR changes, and user preference in one round of schedule--execute iterations.

\begin{llmpromptbox}{Scheduling Path Template Example}
Timing (Aggressive) $\rightarrow$ Timing (Incremental) $\rightarrow$ Power $\rightarrow$ Timing (Incremental)
**\textbf{Design summary}**: The design exhibits severe timing violations. After aggressive timing optimization, several violations remain.
We then perform incremental timing optimization with a different algorithm.
After they are fixed, we perform power recovery.
The power-recovery step may introduce timing degradation that requires incremental timing optimization.\\
**\textbf{Target selection}**: Use an aggressive timing pass to gain slack fast, then an incremental timing pass to stabilize, then optimize power, and end with incremental timing cleanup. Switch away from timing if it becomes well-exploited with no gains.\\
**\textbf{Option selection}**: Aggressive timing is **\textbf{exploration-heavy}** (broad option coverage) to expose limits.
Incremental timing is **\textbf{exploitation-heavy}**, using only options shown to help and avoiding prior no-effect attempts. 
Power stage starts with exploration to find which power components can be reduced safely, then exploits the best component-focused options.\\
**\textbf{Guidelines}**: Don’t chase truly non-fixable timing violations. If aggressive timing increases power/area, explicitly recover power afterward (and note whether area needs a later pass). 
After power changes, always run a timing cleanup to repair any regressions while minimizing new side-effects.
\end{llmpromptbox}
\section{Supplementary Experimental Settings}
\subsection{LLM Settings}
The default model for the agents is GPT-5-mini (2025-08-07), with reasoning effort set to low.
DeepSeek-V3.2 (2025-12-01) is used with reasoning turned off, i.e., the deepseek-chat model.
GPT-4o-mini does not have reasoning effort settings.
The temperature is set to 1 and top\_p is also 1 for all models.
\subsection{Design Statistics}
\label{sec:designStats}
We provide the original (pre-optimization) design statistics used in \Cref{tab:asap7-power-breakdown-original}.
The designs suffer from severe timing violations before ECO.
The worst slack usually has the same order of magnitude as the clock period.
\begin{table}[!h]
    \centering
    \small
    \caption{Original ASAP7 design statistics.}
    \label{tab:asap7-power-breakdown-original}
    \begin{tabular}{l|ccccc}
        \hline
        Design & \begin{tabular}[c]{@{}c@{}}Clock\\Period (ps)\end{tabular} & \begin{tabular}[c]{@{}c@{}}Total\\Slack (ps)\end{tabular} & \begin{tabular}[c]{@{}c@{}}Worst\\Slack (ps)\end{tabular} & Area ($\mu m^2$) & Power (W) \\
        \hline
        NVDLA\_partition\_m & 4e+02 & -1.909e+04 & -3.657e+02 & 2.728e+03 & 5.250e-03 \\
        NVDLA\_partition\_p & 4e+02 & -2.073e+05 & -2.784e+02 & 9.570e+03 & 5.450e-02 \\
        ariane136 & 7e+02 & -2.189e+05 & -1.667e+02 & 3.266e+04 & 5.928e-01 \\
        mempool\_tile\_wrap & 2.6e+03 & -2.133e+06 & -4.777e+02 & 2.813e+04 & 4.640e-02 \\
        aes\_256 & 5.5e+02 & -5.774e+04 & -5.820e+01 & 2.728e+04 & 3.037e-01 \\
        hidden1 & 7e+02 & -4.237e+04 & -1.314e+02 & 5.355e+03 & 2.200e-02 \\
        hidden2 & 7e+02 & -1.325e+05 & -1.823e+02 & 3.307e+04 & 5.863e-01 \\
        hidden5 & 3e+02 & -1.512e+05 & -8.691e+01 & 2.734e+04 & 4.440e-01 \\
        \hline
    \end{tabular}
\end{table}

We further provide detailed breakdown on power components in the dataset.
The power consumption can be mainly attributed to clock power (Clock), power on registers (Reg), and power on combinational gates (Comb).
The clock power group does not include leakage power because the original ICCAD'24 dataset~\cite{PD-ICCAD24-Wu} does not contain a clock tree.
Hence, clock power can be several orders of magnitude larger than register and combinational-gate power.
\begin{table}[!h]
    \centering
    \small
    \caption{Original ASAP7 design power breakdown.}
    \label{tab:asap7-stats-original}
\resizebox{\linewidth}{!}{%
\begin{tabular}{l|ccc|cc|ccc|ccc}
  \hline
  Design & \multicolumn{3}{c|}{Total} & \multicolumn{2}{c|}{Clock} & \multicolumn{3}{c|}{Reg} & \multicolumn{3}{c}{Comb} \\
  & Total & Leak & Dyn & Total & Dyn & Total & Leak & Dyn & Total & Leak & Dyn \\
  \hline
  NVDLA\_partition\_m & 5.253e-03 & 3.417e-06 & 5.250e-03 & 3.999e-03 & 3.999e-03 & 2.406e-04 & 4.005e-07 & 2.402e-04 & 9.286e-04 & 3.016e-06 & 9.256e-04 \\
  NVDLA\_partition\_p & 5.450e-02 & 9.041e-06 & 5.450e-02 & 3.700e-02 & 3.700e-02 & 1.163e-03 & 3.309e-06 & 1.159e-03 & 1.500e-02 & 5.733e-06 & 1.501e-02 \\
  aes\_256 & 3.037e-01 & 2.861e-05 & 3.037e-01 & 2.750e-02 & 2.750e-02 & 4.110e-02 & 4.525e-06 & 4.110e-02 & 2.351e-01 & 2.409e-05 & 2.351e-01 \\
  ariane136 & 5.928e-01 & 1.760e-02 & 5.753e-01 & 0.000e+00 & 0.000e+00 & 0.000e+00 & 0.000e+00 & 0.000e+00 & 4.150e-02 & 2.091e-05 & 4.143e-02 \\
  hidden1 & 2.200e-02 & 4.336e-06 & 2.200e-02 & 1.900e-02 & 1.900e-02 & 4.623e-04 & 2.333e-06 & 4.599e-04 & 2.151e-03 & 2.002e-06 & 2.150e-03 \\
  hidden2 & 5.863e-01 & 1.720e-02 & 5.692e-01 & 0.000e+00 & 0.000e+00 & 0.000e+00 & 0.000e+00 & 0.000e+00 & 4.780e-02 & 2.134e-05 & 4.780e-02 \\
  hidden5 & 4.440e-01 & 3.492e-05 & 4.440e-01 & 3.730e-02 & 3.730e-02 & 6.070e-02 & 3.600e-06 & 6.070e-02 & 3.460e-01 & 3.132e-05 & 3.459e-01 \\
  mempool\_tile\_wrap & 4.640e-02 & 2.610e-03 & 4.380e-02 & 4.816e-03 & 4.816e-03 & 5.211e-04 & 2.447e-06 & 5.186e-04 & 1.620e-02 & 2.930e-05 & 1.618e-02 \\
  \hline
\end{tabular}
}

\end{table}

\subsection{RL Algorithm Settings}
\label{subsec:RL-setup}
We adapt A3C~\cite{Arch-AAAI24-Bai} and PPO~\cite{PD-TODAES26-Hsiao} as baselines, and the RL training settings are shown in \Cref{tab:appendixRL}.
For training, they are trained with 50 episodes.
Each episode launches one optimization.
Thus, RL receives more QoR feedback signals than our agentic framework.
We give a large reward for total slack since timing is the top priority during chip optimization.
We also give additional bonus when timing violations are eliminated.
\begin{table}[h!]
\centering
\small
\begin{tabular}{|l|l|l|}
\hline
Category & Setting & Value  \\
\hline
Training & episode & 50\\
Training & num-iteration & 10 \\
Training & gamma & 0.99 (discount factor) \\
Training & lambda & 0.95 (GAE smoothing for advantages) \\
Training & learning-rate & 1e-3 \\
Training & clip-grad-norm & 0.5 \\
Training & beta & 0.5 (critic loss mixing weight) \\
Training & alpha & 0.01 (entropy weight for exploration) \\
Reward & Reward weights & total slack=10.0, worst slack=3.0, power=3.0, area=3.0 \\
Reward & Penalties/bonuses & violation\_fixed\_bonus=10 \\
\hline
\end{tabular}
\label{tab:appendixRL}
\end{table}

\section{Additional Experimental Results with Customized Optimization Tasks}
\label{sec:power-breakdown}
The power breakdown of the original designs is reported in \Cref{tab:asap7-power-breakdown-original}.
Because clock-network power dominates total power and is not the primary focus of the ICCAD'24 benchmark~\cite{PD-ICCAD24-Wu}, we evaluate power consumption only for combinational cells and registers.
For area, since chip area is determined by the final layout rather than individual cells, we do not perform a breakdown analysis.
For customized optimization, we allow up to a 5\% degradation in worst slack relative to the clock period.

Post-optimization power results for combinational cells and registers under the default (``Ours'') and customized (``Custom'') objectives are shown in \Cref{tab:custom_comb_power_components} and \Cref{tab:custom_reg_power_components}, respectively.
Because gate sizing and buffer insertion/removal primarily affect combinational logic, the customized objective yields a more pronounced reduction in combinational power compared to total-power trends in the main text.

For register power, the ariane136 and hidden2 designs exhibit zero reported register power and are therefore excluded from comparison.
This behavior arises from the use of asynchronous flip-flops, which are not categorized as standard registers.
Although sizing and buffering rarely target registers directly, the customized objective still achieves register power reductions in the remaining designs.

Overall, these results demonstrate that the natural-language understanding capability of LLM agents enables accurate interpretation of user-defined objectives and effective scheduling of chip optimizations to satisfy customized trade-offs.
\begin{table}[!h]
\caption{Combinational power components comparison. Reduction is computed as $(\text{Ours} - \text{Custom})/\text{Ours} \times 100\%$.}
    \centering
    \footnotesize
    \setlength{\tabcolsep}{2pt}
    \begin{tabular}{l c c c c c c c c c}
    \hline
    Design & \multicolumn{3}{c}{Total (W)} & \multicolumn{3}{c}{Leak (W)} & \multicolumn{3}{c}{Dyn (W)} \\
     & Ours & Custom & Reduction & Ours & Custom & Reduction & Ours & Custom & Reduction \\
    \hline
    aes\_256 & 2.39e-1 & 2.32e-1 & 2.9\% & 1.34e-4 & 1.07e-4 & 20.1\% & 2.39e-1 & 2.32e-1 & 2.9\% \\
    ariane136 & 4.16e-2 & 3.29e-2 & 20.9\% & 4.31e-5 & 1.74e-5 & 59.6\% & 4.15e-2 & 3.29e-2 & 20.7\% \\
    hidden1 & 2.22e-3 & 2.14e-3 & 3.6\% & 1.95e-5 & 1.07e-5 & 45.1\% & 2.20e-3 & 2.13e-3 & 3.2\% \\
    hidden2 & 3.68e-2 & 3.63e-2 & 1.4\% & 1.32e-5 & 1.25e-5 & 5.3\% & 3.68e-2 & 3.63e-2 & 1.4\% \\
    mempool\_tile\_wrap & 1.80e-2 & 1.74e-2 & 3.3\% & 1.30e-3 & 1.13e-3 & 13.1\% & 1.68e-2 & 1.62e-2 & 3.6\% \\
    \hline
    \end{tabular}
    \label{tab:custom_comb_power_components}
\end{table}

\begin{table}[!h]
\caption{Register power components comparison. Reduction is computed as $(\text{Ours} - \text{Custom})/\text{Ours} \times 100\%$; ``--'' denotes non-comparable power components.}
    \centering
    \footnotesize
    \setlength{\tabcolsep}{2pt}
    \begin{tabular}{l c c c c c c c c c}
    \hline
    Design & \multicolumn{3}{c}{Total (W)} & \multicolumn{3}{c}{Leak (W)} & \multicolumn{3}{c}{Dyn (W)} \\
     & Ours & Custom & Reduction & Ours & Custom & Reduction & Ours & Custom & Reduction \\
    \hline
    aes\_256 & 4.23e-2 & 4.10e-2 & 3.1\% & 4.53e-6 & 4.53e-6 & 0.0\% & 4.23e-2 & 4.10e-2 & 3.1\% \\
    ariane136 & 0 & 0 & -- & 0 & 0 & -- & 0 & 0 & -- \\
    hidden1 & 4.63e-4 & 4.61e-4 & 0.4\% & 2.33e-6 & 2.33e-6 & 0.0\% & 4.60e-4 & 4.59e-4 & 0.2\% \\
    hidden2 & 0 & 0 & -- & 0 & 0 & -- & 0 & 0 & -- \\
    mempool\_tile\_wrap & 5.25e-4 & 5.15e-4 & 1.9\% & 2.45e-6 & 2.39e-6 & 2.4\% & 5.23e-4 & 5.13e-4 & 1.9\% \\
    \hline
    \end{tabular}
    \label{tab:custom_reg_power_components}
\end{table}

\section{Prompt Examples}
In this section, we provide the prompts used in our framework.
Some detailed descriptions are omitted.
For detailed contents related to the confidential manual of the PrimeECO tool, we provide mock contents.
A reader with access to the tool license should be able to easily complement them.
\subsection{Prompts of Schedulers}
\begin{llmpromptbox}{Schedulers Generating Summary and Query}
\textcolor{myred}{\textbf{System}}: You are an expert IC design ECO (Engineering Change Order) optimization engineer responsible for comprehensive analysis and scheduling.
You will work as a reviewer engineer to evaluate the current design state, optimization history, and unfixable reasons regarding timing, power, and area. Then generate a single query to retrieve relevant knowledge for both optimization target selection and option selection strategy.

**\textbf{ECO Background}**: ECO is an incremental design optimization process that iteratively improves the design by fixing violations and optimizing metrics such as timing, area, and power. Each iteration involves analyzing the current design state and optimization histories and the optimization of one target metric with specific optimization options...

**\textbf{Report Content}**:
- CURRENT DESIGN STATE: The current design state includes timing, power, and area metrics after the most recent optimization iteration.
- OPTIMIZATION HISTORY: The optimization history includes the design states and already performed optimization commands from all previous iterations.
- UNFIXABLE ISSUES HISTORY: The unfixable issues history includes the design states and reasons for unfixable violations from all previous iterations.
- OBJECTIVES: The objectives describe optimization priorities for this run.

**\textbf{Task}**: Based on your analysis, generate the following content (Do not repeat design states, history, or unfixable reasons in your evaluation with detailed values.):
1. Your evaluation on the optimization trend of timing, power, area from previous iterations.
2. Your evaluation on unfixable reasons of timing, power, area from previous iterations.
3. Generate exactly 1 query to retrieve relevant knowledge to assist in decision-making. Focus on:
- Optimization target selection strategy based on your trend evaluation.
- Optimization option selection strategy (Exploration/Exploitation) based on unfixable reasons.

**\textbf{Evaluation Content}**:
1. Evaluate the trend of timing, area, power metrics. Use the best achieved value (the default is iteration 0) as the optimization baseline.
2. Evaluate the spent optimization iterations on optimizing all three targets (timing, area, power). The iteration budget at iteration 0 will be the total budget.
3. Evaluate the iterations that they have major optimization effect, e.g., improving timing at the cost of increasing power and area.\\
...
**\textbf{Evaluation Guidelines}**: 
1. Only make evaluation on given data, do not make any conjecture and comments on optimization strategy or priorities.
2. The command at each iteration will affect the design state at the next iteration, not the current one.\\
...

**\textbf{RESPONSE FORMAT}**: Always respond in valid JSON format.
Example:
\{
  "evaluation": "Your evaluation here, no more than 200 words.",
  "query": "query on combined target and option selection strategy, use plain text, no more than 40 words."
\}

\medskip
<\textcolor{myyellow}{\textbf{User}}>: 
CURRENT DESIGN STATE: The same as optimization history, except the history keeps lists of those metrics.
OPTIMIZATION HISTORY: 
"Iterations :",
"- Setup violations:  paths",
"- Hold violations: paths",
"- Setup critical slack:  ps",
"- Hold critical slack:  ps",
"- Setup total negative slack:  ps",
"- Hold total negative slack:  ps",
"- Total power: W",
"- Dynamic power: W",
"- Leakage power: W",
"- Design area:  $um^2$",
"- Executed commands: "
UNFIXABLE REASON HISTORY: ...
Now generate the response in the required JSON format.

\medskip
<\textcolor{mylight}{\textbf{Agent}}>:
\{'evaluation': 'Baseline only (iteration 0). Setup timing is severely degraded (many violations, large negative critical and total negative slack); hold is clean. Power and area are baseline values with no prior changes...',
 'query': 'Given no prior ECO runs, nine iterations left, and objective to fix heavy setup violations with minimal power/area impact...'\}
\end{llmpromptbox}

\begin{llmpromptbox}{Schedulers Selecting Optimization  Objective}
\textcolor{myred}{\textbf{System}}: You are an expert IC design ECO (Engineering Change Order) optimization engineer responsible for optimization scheduling.

**\textbf{TASK}**: You will work as a scheduler to select the optimization target and optimization option strategy for the current ECO iteration based on (with Priority):
1. The evaluation on optimization trends and unfixable reasons, 
2. The reflection over strategy.
3. The template target/option selection strategy (you can refer to them but override them with your situation).
You need to schedule as follows:
1. Select the optimization target for the current iteration (timing/power/area) and choose the optimization selection strategy (Exploration/Exploitation).
2. Summarize your analysis process for final decision making.

**\textbf{ECO Background}**: ECO is an incremental design optimization process that iteratively improves the design by fixing violations and optimizing metrics such as timing, area, and power. 

**\textbf{Report Content}**:
- CURRENT DESIGN STATE: The current design state includes timing, power, and area metrics after the most recent optimization iteration.
- OPTIMIZATION HISTORY: The optimization history includes the design states and already performed optimization commands from all previous iterations.
- UNFIXABLE ISSUES HISTORY: The unfixable issues history includes the design states and reasons for unfixable violations from all previous iterations.
- OBJECTIVES: The objectives describe optimization priorities for this run.

**\textbf{GUIDELINES}**:
1. Please combine the evaluation and reference strategy for final decision.
2. For detailed optimization trends and unfixable reasons that are not clear in the evaluation, you can refer to the detailed attached report.
**\textbf{RESPONSE FORMAT}**: Always respond in valid JSON format.
...

\medskip
<\textcolor{myyellow}{\textbf{User}}>: 
CURRENT DESIGN STATE: The same as optimization history, except the history keeps lists of those metrics.
OPTIMIZATION HISTORY: 
"Iterations :",
"- Setup violations:  paths",
"- Hold violations: paths",
"- Setup critical slack:  ps",
"- Hold critical slack:  ps",
"- Setup total negative slack:  ps",
"- Hold total negative slack:  ps",
"- Total power: W",
"- Dynamic power: W",
"- Leakage power: W",
"- Design area:  $um^2$",
"- Executed commands: "
UNFIXABLE REASON HISTORY: ...

Evaluation on ECO target and option selection:
\$ evaluation,
Reflection on Strategy:
\$ reflection,
Template Strategy on ECO target and option selection (Could be overridden):
\$ rag\_content,
Now generate the response in the required JSON format.

\medskip
<\textcolor{mylight}{\textbf{Agent}}>:
\{'target': 'timing',
 'option': 'Exploration',
 'reasoning': 'Setup timing is severely degraded (many setup violations, large negative WNS/TNS) while hold is clean and no prior ECO history or unfixable reasons exist. Given the objective to recover timing with minimal power/area overhead, the first iteration should aggressively focus on setup timing. Start with an exploration strategy to try diverse timing-recovery options, reveal easy wins and surface any unfixable causes, and avoid prematurely committing to high-impact fixes that might incur unnecessary power/area cost. Exploration preserves ability to exploit the best options in subsequent iterations and keeps runtime budget flexible for targeted follow-ups.'\}
\end{llmpromptbox}

\subsection{Prompts of Executor Agents}

\begin{llmpromptbox}{Executor Generating Evaluation for RAG}
\textcolor{myred}{\textbf{System}}: **\textbf{Background}**: You are an expert in multi-iteration Engineering Change Order (ECO) for IC design with limited iteration budget. You will be given optimization trends, including current design state, optimization history, and unfixable issues. You will be also given the optimization strategy and objectives. You will generate evaluation on current timing optimization trends and unfixable reasons to help making informed decisions.

**\textbf{Report Content}**:
- CURRENT DESIGN STATE: The current design state includes timing, power, and area metrics after the most recent optimization iteration.
- OPTIMIZATION HISTORY: The optimization history includes the design states and already performed optimization commands from all previous iterations.
- UNFIXABLE ISSUES HISTORY: The unfixable issues history includes the design states and reasons for unfixable violations from all previous iterations.
- OBJECTIVES: The design objectives for the ECO run, including timing, power, and area goals.

**\textbf{Task}**: Based on your analysis, generate your evaluation of the timing, power, and area trends from previous iterations. Do not repeat detailed values from reports in the evaluation.
**\textbf{Evaluation Format}**: Given [your evaluation in plain text], what timing-fix guidelines and command options should be used?

Evaluation Content (You do not need to cover all of them. You can only cover the ones you found in reports):
1. Evaluate the trend of timing. Use the best achieved value (the default is iteration 0) as the optimization baseline.
2. Evaluate the current existing unfixable reasons.\\
...
Evaluation Guidelines:
1. Provide separate evaluations for setup and hold violations. For unfixable reasons, identify whether they belong to setup or hold from the command executed at that iteration.
2. When evaluating trends, mind the trade-off between timing, power, and area across optimization iterations.\\
...

DEFINITIONS OF UNFIX REASONS and potential solutions:...
RESPONSE FORMAT: Always respond in plain text.

\medskip
<\textcolor{myyellow}{\textbf{User}}>: 
CURRENT DESIGN STATE: The same as optimization history, except the history keeps lists of those metrics.
OPTIMIZATION HISTORY: 
"Iterations :",
"- Setup violations:  paths",
"- Hold violations: paths",
"- Setup critical slack:  ps",
"- Hold critical slack:  ps",
"- Setup total negative slack:  ps",
"- Hold total negative slack:  ps",
"- Total power: W",
"- Dynamic power: W",
"- Leakage power: W",
"- Design area:  $um^2$",
"- Executed commands: "
UNFIXABLE REASON HISTORY: ...

Evaluation on ECO target and option selection:\\
\$evaluation\\
Reflection on Strategy:\\
\$reflection\\
Template Strategy on ECO target and option selection:\\
\$rag\_content\\

\medskip
<\textcolor{mylight}{\textbf{Agent}}>:
Given: Setup dominates with large WNS/TNS;  no prior fixes or unfixable reasons identified; no trade-off observed yet. Exploration should prioritize setup fixes with area/power degradation, starting with high-effort...
Check potential command for set up fixing...
\end{llmpromptbox}

\begin{llmpromptbox}{Executor Generating EDA Tool Command}
\textcolor{myred}{\textbf{System}}: You are an expert in IC design timing fixing responsible for fixing timing violations.
Your task is to analyze the current timing violations and generate an effective command based on: 1. optimization strategies and objectives 2. The current design state and optimization history. 3. The previous unfixable analysis (if any).
Then generate an optimal optimize timing command.

COMMAND FORMAT:...

DEFINITIONS OF UNFIX REASONS:...

RESPONSE FORMAT: Always respond in valid JSON format.

<\textcolor{myyellow}{\textbf{User}}>:\\
Strategy: \$strategy\\
Objectives: \$objectives\\
Guidelines: \$rag\_retrieved\_content\\
REFLECTION on command usage: \$ reflection\\
CURRENT DESIGN STATE:...
OPTIMIZATION HISTORY: ...
UNFIXABLE REASON HISTORY: ...\\
Now generate the command.

\medskip
<\textcolor{mylight}{\textbf{Agent}}>:
timing\_opt -methods gate\_sizing -area\_overhead low -slack\_effort high ...
\end{llmpromptbox}

\subsection{Prompts of Reflectors}
\begin{llmpromptbox}{Schedule-level Reflection}
<\textcolor{myred}{\textbf{System}}>: 
You are an expert IC design ECO (Engineering Change Order) optimization engineer responsible for optimization history analysis.
You will work as a reviewer engineer to evaluate the optimization history regarding timing, power, and area w.r.t the Objectives. The response should be less than 100 words.

**\textbf{ECO Background}**: ECO is an incremental design optimization process that iteratively improves the design by fixing violations and optimizing metrics such as timing, area, and power. Each iteration involves analyzing the current design state and optimization histories and the optimization of one target metric with specific optimization options...

**\textbf{Response Guidelines}**:

Produce a comment on optimization target ordering and effort selection Pareto frontier observations. Provide guidance only; no tool commands or lists.
Each trace consists of a sequence of commands with opt\_[timing/power/area] and metrics change.
Focus on how objectives were selected and how many iterations were spent on each target.
Comment on whether some targets need more or fewer iterations for optimization.
Highlight whether any target needs recovery after being sacrificed when optimizing another target.

\medskip
<\textcolor{myyellow}{\textbf{User}}>:
Objectives:
Optimize timing to fix TNS and WNS violations, with minimal power and area overheads.

Pareto optimal traces:
Trace 1: commands: [opt\_timing, opt\_timing, ...]; timing:[...], power: [...], area: [...]
Trace 2:...

\medskip
<\textcolor{mylight}{\textbf{Agent}}>:
Traces show heavy bias to timing fixes (many successive timing iterations) with occasional power/area tweaks. Pareto frontier candidates that interleave modest power/area effort achieve lower power and area with acceptable timing; one trace shows severe timing regression after aggressive power moves indicating unsafe tradeoff. Recommendation: front-load targeted timing passes until obvious slack improvement stalls, then apply focused power/area recovery to remove overheads. Avoid repeated blind timing iterations with diminishing returns; use limited, measured power/area iterations to recover any metric sacrificed by prior timing optimizations.
\end{llmpromptbox}

\begin{llmpromptbox}{Execute-level Reflection}
<\textcolor{myred}{\textbf{System}}>: 
You are an expert IC design ECO (Engineering Change Order) optimization engineer responsible for comprehensive analysis.
You will work as a reviewer engineer to evaluate the optimization history regarding timing, power, and area w.r.t the Objectives. The response should be less than 100 words.

**\textbf{ECO Background}**: ECO is an incremental design optimization process that iteratively improves the design by fixing violations and optimizing metrics such as timing, area, and power. Each iteration involves analyzing the current design state and optimization histories and the optimization of one target metric with specific optimization options.

**\textbf{Response Guidelines}**:

You should analyze the trend of timing, power, area change in iterations related to objectives and the trend of unfixable reasons' change. 
You should comment on the effectiveness of command that optimize the same objective with different options.
You should comment on command options relate to unfixable reasons and whether those options helped.

\medskip
<\textcolor{myyellow}{\textbf{User}}>:
Objectives:
Optimize timing to fix TNS and WNS violations, with minimal power and area overheads.

Pareto optimal traces:
Trace 1: commands: [opt\_timing, opt\_timing, ...]; timing:[...], power: [...], area: [...]
Trace 2:...

\medskip
<\textcolor{mylight}{\textbf{Agent}}>:
All traces quickly resolved WNS from -58ps to 0-1 by early iterations with minimal power (+0.001-0.005) and small area changes; power/area commands (gate\_sizing, buffer\_removal) further reduced power/area in traces 2-3 and 6. Using -area\_overhead  enabled access to larger/blocked cells. Slack efforts focused effort effectively but over‑aggressive use (trace 3) caused later timing regressions (new negative WNS/TNS) — indicates tradeoff misuse.
Overall commands targeting same objective with different options were effective when combined judiciously; avoid aggressive slack narrowing without power/area recovery steps.
\end{llmpromptbox}

\end{document}